\documentclass[aps,preprint]{revtex4}%
\usepackage{amsmath}
\usepackage{amssymb}%
\usepackage{amsfonts}%
\usepackage{graphicx}

\begin{document}
\title[Vacuum forces in MEMS]{The Role of Quantum Vacuum Forces in\\Microelectromechanical Systems }
\author{G. Jordan Maclay}
\affiliation{Quantum Fields LLC, Richland Center WI 53581 USA}
\email{jordanmaclay@quantumfields.com}

\begin{abstract}
The presence of boundary surfaces in the vacuum alters the ground state of the
quantized electromagnetic field and can lead to the appearance of vacuum
forces. \ In the last decade, landmark measurements of the vacuum stress
between conducting uncharged parallel plates. Recently the first micromachined
MEMS (microelectromechanical system) device was fabricated that utilizes the
Casimir force between parallel plates. The $1/d^{4}$ force dependence allows
the device to serve as a highly sensitive position sensor. \ The are many
other examples of quantum vacuum forces and effects besides the well known
parallel plate Casimir force. \ Here we discuss potential roles of quantum
vacuum forces and effects in MEMS systems and other systems. \ With the
growing capability in nanofabrication, some of the roles may be actualized in
the future. \ Because of the computational complexity, no theoretical results
are yet available for a number of potentially interesting geometries and we
can only speculate.

\end{abstract}
\date{July 28, 2006}
\startpage{1}
\endpage{2}
\maketitle

\section{Introduction}

Zero-point field energy density is a simple and inexorable property of a
quantum field, such as the electromagnetic field, which is a representation of
the Lorentz group of transformations of special relativity. \ For a quantum
field, the canonical position and momentum variables do not commute and
consequently the lowest state of the field has a non-zero energy, just as the
ground state of a quantum mechanical harmonic oscillator is non-zero. \ \ For
the electromagnetic field, if we assume the shortest wavelength photon to be
included in the ground state spectrum has the Planck length of 10$^{-35}$ m,
then the predicted quantum vacuum energy density is enormous, about $10^{114}$
J/m$^{3}$ or, in terms of mass, $10^{95}$ g/cm$^{3}.$ \ Such an enormous
energy density is clearly a puzzling embarrassment to physicists, who for
years routinely discarded this nearly infinite result in renormalization procedures.

However, there are measurable consequences of the zero point energy which
arise because the ground state vacuum electromagnetic field has to meet the
usual boundary conditions for the electromagnetic field. \ It is the effect of
boundaries on the vacuum field that leads to the appearance of vacuum
stresses, so called Casimir forces.

The term \textquotedblright Casimir force\textquotedblright\ $F$ most commonly
refers to the attractive vacuum force per area (pressure) that exists between
two parallel, infinite, uncharged, perfectly conducting plates separated by a
distance $d$ \cite{plunienreview}\cite{bordag}%
\begin{equation}
F=-\frac{K}{d^{4}}%
\end{equation}

where $K=\frac{\pi^{2}\hslash c}{240}.$ \ This force arises from the change in
energy density $E_{pp}$ from the free field vacuum density that occurs between
the parallel plates:%

\begin{equation}
E_{pp}(d)=\frac{K}{3d^{3}}%
\end{equation}

This attractive force $F$\ ,which is normal to the surface, arises because the
surfaces change the mode distribution of the ground state quantized
electromagnetic field. \ In the region between two parallel perfectly
conducting plates, no modes with wavelengths larger than twice the separation
can exist. \ We can also view this force as arising from radiation pressure,
the transfer of momentum from the vacuum to the
surfaces\cite{milonniradpressure}. \ The Casimir effect was first predicted in
1948, but was not measured accurately until the last few
years\cite{mohideenroy}\cite{lamoroux}. \ Corrections for finite conductivity
and surface roughness have been developed for the parallel plate geometry,
\ and the agreement between theory and experiment is now at the 1\% level or
better for separations of about \ 0.1-0.7 $\mu m$ \cite{realplatesmoh}. \ In
actual practice, the measurements are most commonly made with one surface
curved and the other surface flat, and using the proximity force theorem to
account for the curvature. \ This experimental approach eliminates the
difficulties of trying to maintain parallelism at submicron separations.
\ Mohideen and collaborators have made the most accurate measurements to date
in this manner, using an AFM (Atomic Force Microscope) that has a metallized
sphere about 250 $\mu m$ in diameter attached to the end of a cantilever about
200 $\mu m$ long, capable of measuring picoNewton forces. \ The deflection of
the sphere is measured optically as it is moved close to a flat metallized
surface\cite{mohideenroy}. \ The more difficult measurement between two
parallel plates has been made and shown to give results that are consistent
with theory\cite{bressi}. \ Measurements of the force between two parallel
surfaces each with a small (1 nm) sinusoidal modulation in surface height,
have showed that there is a lateral force as well as the usual normal force
when the modulations of the opposing surfaces are not in phase
\cite{chenlateral}. \ Recent measurements have confirmed the predictions,
including effects of finite conductivity, surface roughness, and temperature,
uncertainty in dielectric functions, to the 1-2\% level for the range from
65-300 nm\cite{mohideen2}. \ There is a small uncertainty in the temperature
corrections, particularly at low temperatures \cite{temp}.

\ Casimir forces occur for all quantum fields and can arise from the presence
of surfaces as well as choices of topology of the space.\ \ Initially Casimir
forces for plane surfaces were obtained by computing the change in the vacuum
energy with position. \ Two decades after Casimir's initial predictions, a
method was developed to compute the Casimir force in terms of the local
stress-energy tensor using quantum electrodynamics\cite{brownmaclay}. \ Many
innovations have followed. \ Several approaches to computing electromagnetic
Casimir forces have been developed that are not based on the zero point vacuum
fluctuations directly. \ These approaches appeal to scientists who are
uncomfortable with the quantum electrodynamical model of energy in empty
space. \ In the special case of the vacuum electromagnetic field with
dielectric or conductive boundaries, various approaches suggest that Casimir
forces can be regarded as macroscopic manifestations of many-body retarded van
der Waals forces, at least in simple geometries with isolated atoms\cite{pwm},
\cite{Power}. \ Casimir effects have also been derived and interpreted in
terms of \textit{source} fields in both conventional \cite{pwm} and
unconventional \cite{schw} quantum electrodynamics, in which the fluctuations
appear within materials instead of outside of the materials. \ Lifshitz
provided a detailed computation of the Casimir force between planar surfaces
by assuming that stochastic fluctuations occur in the tails of the
wavefunctions of atoms that leak into the regions outside the surface, and can
lead to induced dipole moments in atoms in a nearby surface, which leads to an
a net retarded dipole-induced dipole force between the planar
surfaces\cite{lifshitz}. \ These various approaches differ in how they
visualize the fluctuations of the electromagnetic field, but give consistent
results in the few cases of simple geometries which have been
computed\cite{maclayandmilonni}. \ It may be that these diverse approaches
will display differences for computation of geometries with curvature, or for
computations of the forces between separated, curved
surfaces\cite{maclaysqueezed}.

Parallel plate Casimir forces go inversely as the fourth power of the
separation between the plates. \ The Casimir force per unit area $F$ between
perfectly conducting plates is equivalent to about 1 atm pressure\ at a
separation of 10 nm, and so is a candidate for actuation of MEMS
(MicroElectroMechanical Systems).\ In MEMS, surfaces may come into close
proximity with each other, particularly during processes of etching
sacrificial layers in the fabrication process. \ In 1995 the first analysis of
a dynamic MEMS structure that used vacuum forces was presented by Serry et
al\cite{serry}. \ They consider an idealized MEMS component resembling the
original Casimir model of two parallel plates, except that one of the plates
is connected to a stationary surface by a linear restoring force and can move
along the direction normal to the plate surfaces. \ The Casimir force between
the two plates, together with the restoring force acting on the moveable
plate, results in a bistable system with two equilibrium separtions. \ The
larger separation is a stable equilibrium and the smaller one is unstable,
leading to the collapse of the movable surface into stationary plate. The
analysis demonstrates that the Casimir effect could be used to actuate a
switch, and might be responsible in part for the \textquotedblleft stiction"
phenomenon in which micromachined membranes are found to latch onto nearby
surfaces. \ If the movable surface is vibrating, then an \textquotedblleft
anharmonic Casimir oscillator" (ACO) results.

To explore stiction in common MEMS configurations, Serry et al computed the
deflection of membrane strips and the conditions underwhich they would
collapse into nearby surfaces\cite{serry2}. \ Measurements were done by Buks
et al on cantilever beams to investigate the role of Casimir forces in
stiction\cite{buk}. \ An experimental realization of the ACO in a
nanometer-scale MEMS system was recently reported by Chan et al \cite{chan}.
In this experiment the Casimir attraction between a 500 $\mu$m-square plate
suspended by torsional rods and a gold-coated sphere of radius 100 $\mu$m was
observed as a sharp increase in the tilt angle of the plate as the
sphere-plate separation was reduced from 300 nm to 75.7 nm. \ This
\textquotedblleft quantum mechanical actuation" of the plate suggests
\textquotedblleft new possibilities for novel actuation schemes in MEMS based
on the Casimir force" \cite{chan}. \ In a refinement of this experiment, a
novel proximity sensor was demonstrated in which the plate was slightly
oscillated with an AC signal, and the deflection amplitude observed gave an
indication of the precise location of the nearby sphere\cite{chan2}. \ A
measurement using a similar torsion oscillator was recently reported using
gold on the sphere and chromium on the plate\cite{decca}.

MEMS currently employed in sensor and actuator technology generally have
component separations on the order of microns, where Casimir effects are
negligible. \ Smaller distances between MEMS components are desirable in
electrostatic actuation schemes because they permit smaller voltages to be
used to generate larger forces and torques. \ Casimir effects will be of
increasing significant in microelectromechanical systems (MEMS) as further
miniaturization is realized \cite{serry}.

\section{Limitations of Current Theoretical Calculations of Vacuum Forces}

The parallel plate geometry (and the approximately equivalent sphere-plate
geometry or sphere-plate with small deviations geometry) is essentially the
only geometry for which experimental measurements have been conducted and the
only geometry for which the vacuum forces between \textbf{two separate
surfaces} (assumed to be infinite) have been computed. \ \ Vacuum forces are
know to exist in other experimental configurations between separate surfaces,
but rigorous calculations based on QED (quantum electrodynamics) are very
difficult and have yet to be completed\cite{maclaysqueezed}. \ Since it is
experimentally possible to measure forces between various separate surfaces,
with the improvement in experimental techniques, theoreticians may soon see
the need for such computations.

Calculations of vacuum stresses for a variety of geometric shapes, such as
spheres, cylinders, rectangular parallelepipeds, and wedges are reviewed in
\cite{plunienreview}\cite{bordag}\cite{miltonbook}. \ In general, calculations
of vacuum forces become very complex when the surfaces are curved,
particularly with right angles. \ Divergences in energy appear, and there are
disagreements about the proper way to deal with these
divergences\cite{Deutsch}.\ \ The material properties, such as the dielectric
constant and plasma frequency of the metal and the surface roughness also
affect the vacuum forces. \ In addition, in the usual calculations only a
spatial average of the force for a given area for the ground state of the
quantum vacuum field is computed, and material properties, such as binding
energies, are ignored, a procedure which Barton has questioned
recently\cite{bartonboston}\cite{barton2}\cite{maclaypra}.

Computation has shown that the vacuum stress on a spherical metal shell, a
cubical shell, or a solid dielectric ball is a repulsive uniform force, or
directed outward. \ Because of the very special nature of the parallel plate
geometry and the high degree of symmetry of the cube and sphere, it is not
reliable to make generalizations about the behavior of vacuum forces based on
these special geometries. \ The vacuum forces on the faces of conductive
rectangular boxes or cavities show very different features compared to those
of the parallel plate, the cube, and the sphere. \ For a rectangular
parallelepiped cavity, the total force on a given face (the differential force
integrated over the entire face) can be positive, zero, or negative depending
on the ratio of the sides of the box\cite{maclaypra}\cite{ambjorn}%
\cite{carlosrectangular}\cite{lukosz}. \ In fact there are cavities that have
zero force on two sides and a positive or negative force on the remaining
side. \ There are boxes for which the change in the vacuum energy is negative
(or positive) and the forces on some walls are attractive while the forces on
the remaining walls are repulsive. \ Indeed it is difficult to get an
intuitive picture of the meaning of these results.

\ From\ a technological viewpoint, it would be useful to be able to\ generate
repulsive vacuum forces as well as attractive vacuum forces. \ From a
fundamental viewpoint, it is unclear how one can have a repulsive force in
vacuum if the force can be correctly modeled as a dipole-induced dipole force.
\ Thus there is great interest in measuring the vacuum forces in different
geometries that are predicted to be repulsive. \ However, there is no easy way
to measure vacuum forces on spheres or rectangular cavities\cite{maclayitamp}.
\ One might consider applying a stress to the spherical shell, and observe the
deformation. \ This is a difficult experiment since the sphere would probably
have to be submicron in diameter for the Casimir force to be large enough to
be measurable. \ Further, the deflection measured would depend on the
properties of the material of which the sphere was made, and such properties
are not included in the usual calculations of the Casimir force\cite{barton2}.
\ Alternatively one might contemplate cutting a sphere in half, and measuring
the force between the two hemispheres using an Atomic Force Microscope.
\ However, the question arises: If we cut a spherical cavity into two
hemispheres, will we find a repulsive force between the two separate surfaces?
\ Or will an attractive force between the edges dominate? No computations have
yet been done for this situation for real materials. \ For optically thin
materials Barton shows the net force will be attractive\cite{bartonboston}%
\cite{barton2}.

Vacuum forces computed for a perfectly conducting cube with thin walls are
also repulsive or outward, and experimentalists have the same conundrum
regarding the meaning of this calculated vacuum force. \ To measure the force
one might imagine freeing one face of the cube, and then moving it very
slightly normal to its surface, in the spirit of the principle of virtual work
dE=-Fdx. \ Unfortunately no one has computed the force between a cube with one
side removed and a nearby surface which is parallel to the missing face. \ We
have attempted to measure the force between an array of open cavities (wall
thickness about 150 nm, cavity width about 200 nm) and a metallized sphere 250
$\mu$m in diameter on an AFM cantilever, and to date have only observed
diminished attractive forces\cite{aiaarepexp}.

Another limitation of the calculations to date for the rectangular cavity, is
that only the total force on each face is computed. \ The differential vacuum
stress in not uniform on each wall, and, in order to avoid issues with
divergences, the differential force is integrated over the face. \ How these
nonuniformities might affect experiments is unknown.

The sign of the Casimir force depends on the magnetic and electric properties
of the materials. \ If it were possible to arbitrarily choose material
properties, repulsive forces could be obtained in a parallel plate geometry,
for example, by choosing one plate to be a perfect conductor ($\epsilon
\rightarrow\infty)$ and the other plate as a perfect magnetic material
($\mu\rightarrow\infty)$ \ for all frequencies, real and imaginary, with a
vacuum in between\cite{kenneth}. \ Other choices have also been suggested,
however, none have been implemented experimentally, and it appears they all
violate fundmental requirements about $\epsilon$ \ and $\mu$ \ for real
materials that arise from causality.conditions\cite{santos}. \ It has been
suggested but not verified that in curved space-time, atoms in certain intense
electric fields may exhibit repulsive forces\cite{pintocurved}.

\section{Vacuum Force Actuated MEMS\ Systems}

We consider a variety of systems whose function is based on existing
calculations of the properties of the ground state of the quantum vacuum.
\ Several different potentially interesting applications are considered in
\cite{maclayandmilonni} \ \ No experimental investigations have yet been
conducted on most of these systems.

\subsection{Structures with Parallel Plates}

We consider the forces and energy balance in several simple structures with:
1. moving parallel plates; 2. moving plates inserted in rectangular cavities,
and 3. pistons moving in rectangular cavities\cite{staif00}.%

\begin{figure}[tbp] \centering
\begin{tabular}
[c]{|l|l|l|}%
{\includegraphics[
trim=0.000000in 0.000000in 3.132293in -0.009578in,
height=0.9081in,
width=1.3629in
]%
{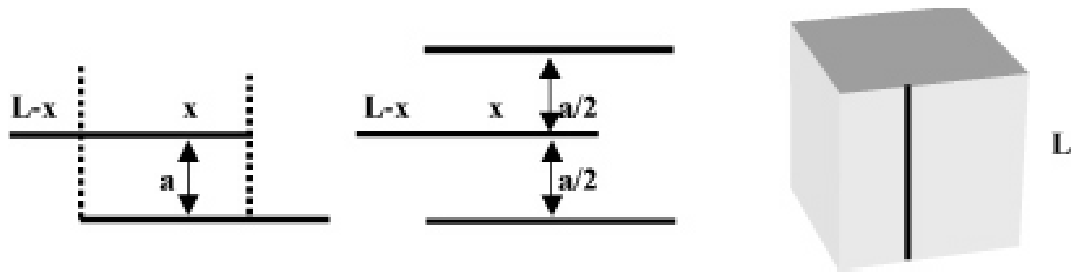}%
}%
&
{\includegraphics[
trim=1.405768in 0.089989in 1.678161in 0.029891in,
height=0.9037in,
width=1.5143in
]%
{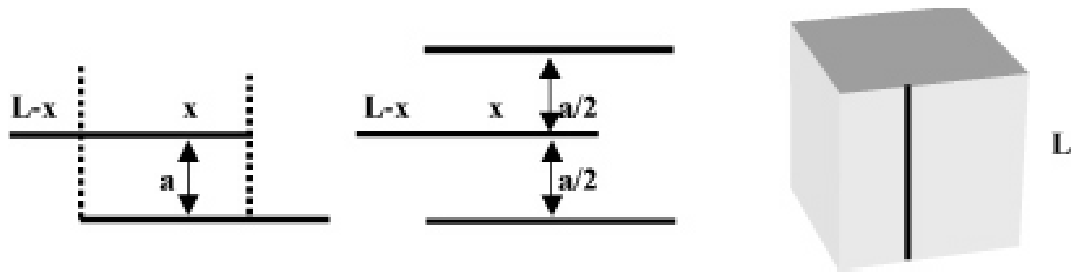}%
}%
&
\raisebox{-0.0104in}{\includegraphics[
trim=2.085300in 0.028908in 0.150600in 0.230221in,
height=0.8968in,
width=0.9461in
]%
{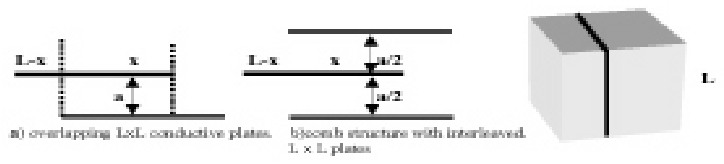}%
}%
\\\hline
\end{tabular}
\caption{ Parallel plates structures with Casimir forces. a) two
overlapping square  (L x L) conductive plates, with lateral or transverse movement
permitted;  b) comb structure of two fixed plates with a third plate between
them which can move laterally, c) conductive moveable piston represented by the black line in a
conductive rectangular cavity L x L x 0.8L.}%
\end{figure}%

\subsubsection{Moving Parallel Plates}

Consider two conducting, overlapping $(x=L)$, square, parallel plates, a
distance $L$ on each side, that are a distance $a$ apart, with $a<<L$. If we
allow the upper plate to approach the lower (fixed) plate quasistatically,
then the attractive force $F_{pp}(a)=-KL^{2}/a^{4}$ does positive mechanical
work during this reversible thermodynamic transformation. We are neglecting
edge effects by assuming that the force is proportional to the area. During
the transformation, the vacuum energy $E_{p}(a)=-KL^{2}/3a^{3}$ between the
plates will be reduced, conserving the total energy in the system. If the
separation decreases from $a_{i}$ to $a_{f}$ , then the energy balance is%
\begin{equation}
E_{p}(a_{f})=E_{p}(a_{i})=-\int_{a_{f}}^{a_{i}}F_{pp}(a)da
\end{equation}

If we then separate the plates quasistatically, letting $a$ increase from
$a_{f}$ to $a_{i}$, we do work on the system to restore it to its initial
state. Over the entire cycle no net work is done, and there is no net change
in the vacuum energy. \ 

Consider an alternative cycle that several investigators have proposed in
order to extract energy from the vacuum fluctuations: After the plates have
reached the point of minimum separation, slide the upper plate laterally until
it no longer is opposite the lower plate $(x=0)$, eliminating the normal
Casimir force, then raise the upper plate to its original height, and slide it
laterally over the lower fixed plate $(x=L)$. \ Finally we allow the plates to
come together as before, extracting energy from the vacuum fluctuations and
doing mechanical work. If no energy were expended in moving the plate
laterally, then this cycle would indeed result in net positive work equal to
the energy extracted from the vacuum. Although no one has yet computed in
detail the lateral forces between offset finite parallel plates, it is highly
probable that such forces are not zero, and that no net extraction of energy
occurs for this cycle. \ We can verify this by a simple approximate
calculation. \ We do know that the vacuum energy is not altered by a single
infinite conducting plate \cite{dewitt}. If we neglect Casimir energy
\textquotedblleft fringing fields,\textquotedblright\ and assume that the
energy density differs from the free field density only in the region in which
the two square $(L\times L)$ plates overlap a distance $x$, where $0<x<L$ (see
Fig. 1a), then we can compute the lateral force $F_{L2}$ between the two
plates using the conservation of energy (principal of virtual work):%
\begin{equation}
F_{L2}(x)=-d[-KLx/3a^{3}]/dx=KL/3a^{3}\label{lat}%
\end{equation}

This constant attractive lateral force tends to increase $x$ or pull the
plates towards each other so they have the maximum amount of overlap. \ In
fact, the positive work done to move one plate laterally a distance $L$
exactly cancels the work extracted from the vacuum fields in moving the plates
from a large separation to a distance $a$ apart, so there is no net change in
total energy (mechanical plus field) in the complete cycle, as expected. \ The
normal Casimir force between these $L\times L$ plates when they are directly
opposite, with complete overlapping ($x=L$), is $L/a$ times larger than the
constant lateral force given by Eq. (\ref{lat}).

\subsubsection{Parallel Plate Comb Structure}

Consider the case of two fixed, square ($L\times L$), parallel plates
separated by a distance a, with a third moveable plate that slides in between
the two parallel plates, separated by a distance $a/2$ from each plate (Fig.
1b). If we neglect vacuum energy \textquotedblleft fringing
fields\textquotedblright, as before, that the vacuum energy is different from
zero only in regions between directly opposing plates, and we can compute the
lateral force on the moveable plate in the middle as minus the derivative of
the vacuum energy. The energy, as a function of the overlap x of the fixed and
moveable plate, is
\begin{equation}
E(x)=-2KLx/3(a/2)^{3}-KL(L-x)/3a^{2}%
\end{equation}

which yields a lateral force $F_{L3}=-dE(x)/dx$ \ equal to%
\begin{equation}
F_{L3}=-5KL/a^{3}%
\end{equation}

This force is $5a/L$ times the normal Casimir force between the plates
separated by a distance $a$, a factor which is typically much less than one.
\ For a device with $a=0.1$ micrometer, $L=1$ mm, the lateral force would be
an easily measurable 31 nanoNewtons. This structure is analogous to the
electrostatic comb drive that is used extensively in MEMS
(microelectromechanical systems) devices. \ One key operational difference
between the Casimir and electrostatic drives is that the Casimir force drive
always yields an inward or attractive force, whereas the voltage on the
electrostatic comb drive can be reversed in polarity, reversing the direction
of the force. Another difference is that the Casimir force comb drive requires
no electrical actuation.

\subsubsection{Inserting Parallel Plates into Rectangular Cavities}

The mechanical behavior of the parallel plate comb configurations is
determined by the negative energy density that arises when $a<<L$. If we
consider cavities that have dimensions in orthogonal directions that are
within about a factor of about 3 of each other, then we can have regions with
positive or negative energy density and can obtain both attractive and
repulsive average forces on sliding plates.

For example, consider a rectangular cavity $L\times L\times a$ formed from
conductive plates. Imagine that the side of length $a$ is constructed so that
we can slowly insert an additional plate (assumed to have zero thickness) in a
direction normal to the $a$ direction, dividing the cavity into two identical
rectangular regions with sides $L\times L\times a/2$. By the conservation of
energy we can compute the average force required to insert this moveable
plate. If we assume that no energy is dissipated within the perfectly
conducting plate during insertion and that the vacuum energy density is
altered only in the region within the cavity as we insert the plate, then the
change in vacuum energy is equal to minus the average force $<F>$ present
during insertion of the movable plate times the distance $L$. Defining
$en(a1,a2,a3)$ as the vacuum energy of a rectangular cavity with sides
$(a1,a2,a3)$, we can express the average force as
\begin{equation}
<F(L,L,a)>=-[2en(L,L,a/2)-en(L,L,a)]/L
\end{equation}

Depending on the ratio of $L/a$ we can obtain positive, zero, or negative
average forces. \ As discussed by \cite{maclaypra}, the energy for a
rectangular cavity is a homogeneous function of the dimensions: $en[\xi a1,\xi
a2,\xi a3]=\xi^{-1}en[a1,a2,a3]$. \ With this information we can evaluate the
average force for several examples. \ Assume $a/L=0.816$, then by numerical
computation we have the final state $en(L,L,0.408L)=0$, and for the initial
state $en(L,L,0.818L)=0.1\hslash c/L$. For this geometry, the mean force is
therefore positive, which means the vacuum field resists the insertion of the
sliding plate:%
\begin{equation}
<F(L,L,0.816L)>=-[0-0.1\hslash c/L]/L=0.1\hslash c/L^{2}.
\end{equation}

For a $L=0.1$ micrometer, the force is about 2.5 picoNewtons, which is near
the current limit of measurability with an AFM. \ Inserting a plane into a
cavity is an interesting operation since it does manifest, at least in theory,
a repulsive force on the movable element. \ Note that we have only computed an
average force during insertion. \ This is consistent with the theoretical
computations which only provide the average vacuum energy density of the cavity.

\subsubsection{Rectangular structures with a moveable piston}

\begin{figure}
[ptb]
\begin{center}
\includegraphics[
trim=0.000000in 4.178802in 0.000000in 0.000000in,
height=2.1958in,
width=2.9308in
]%
{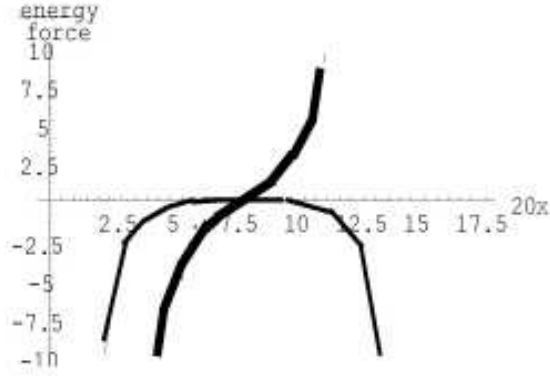}%
\caption{Plot of the vacuum energy (thinner line) and Casimir force (thicker
line) for a 1 x 1 x .8 cavity that is divided into two rectangular cavities by
a sliding piston that moves along the 0.8 direction. Note the variable on the
abscissa is 20x. \ The maximum energy is at abscissa of 8 which corresponds to
the center of the cavity, x = 0.4. For these calculations, we have set
\~{N}c=1 and set L equal to 1 unit. To obtain a numerical result, we use the
MKS value of \~{N}c. If we let L=0.5 micron, then the abscissa is in units of
0.025 micron, and the energy scale is in units of 6.3 x 10-20 joule and the
force scale is in units of 1.26 x 10-13 newton. Forces of this magnitude are
just measurable using AFM technology.}%
\label{box}%
\end{center}
\end{figure}

Consider a rectangular conductive cavity $(L\times L\times a)$ with a
moveable, infinitely thin, perfectly conducting piston that moves along the
$a-$direction, dividing the cavity into two regions, each with its
contribution to the total vacuum energy. We assume the piston is normal to the
$a-$direction (Fig. 1c). We can then numerically compute the total vacuum
energy $E_{P}(L,L,x)$ of the structure as a function of the distance $x$
between the piston and one end of the cavity. From our definition of
$en(a1,a2,a3)$, and the definition $\xi=x/L$ , we have%
\begin{equation}
E_{P}(L,L,x)=[en(1,1,(a/L)-\xi)+en(1,1,\xi)]/L.
\end{equation}
where we have assumed the energies are additive. \ If we differentiate this
equation with respect to $x$, we obtain an expression for the force $F(x)$ due
to the vacuum stresses on the moveable plate. \ Consider an example in which
$a=0.8L$, so $0\leq\xi\leq0.8$. \ Figure 2 shows the dimensionless energy and
force respectively $LEp(L,L,x)$ and $L^{2}F(x)$ as functions of $x $. For
values of $x$ near the center ($x\approx0.4$), the force on the piston is
approximately directly proportional to $x$, and the energy is approximately a
negative parabola with negative curvature. A small deflection from $x=8.0$
leads to a force causing an increased deflection. Thus Figure \ref{box} shows
the state of the system with the piston near the center is unstable: the
piston would be pushed to the closest end of the cavity. More detailed
calculations, in which the divergences were not dropped, have shown that all
divergences exactly cancel for this piston geometry \cite{hertzberg}.%

However, if we include the restoring force on the piston that arises from the
small deflection of a deformable membrane as given by Hooke's Law, then this
configuration might become stable if the material force constant exceeds that
for the Casimir force. \ Of course, once material properties are included,
the\ theoretically computed vacuum forces may be changed. \ In any event,
these results suggest the intriguing possibility of making a structure that
might displays simple harmonic motion for small displacements by employing two
adjacent cubical cavities, with a common face that can be deflected.

\subsection{Vibrating Cavity Walls in MEMS Cavities}

\begin{figure}
[ptb]
\begin{center}
\includegraphics[
height=2.0911in,
width=2.751in
]%
{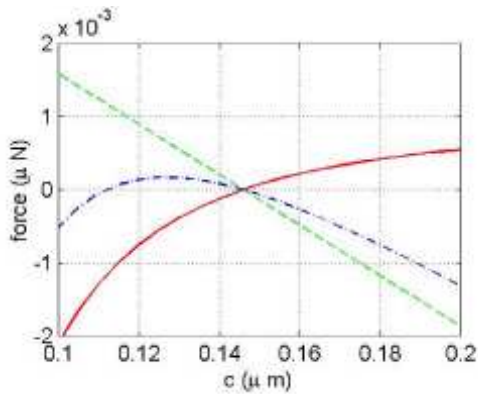}%
\caption{The force on the top surface of a closed, perfectly conducting
rectangular cavity 2 $\mu m$ long by 0.1$\mu m$ wide, as a function of the
depth $c.$ The equilibrium position is c$_{eq}$=0.146 $\mu m.$ \ The dashed
line (- - -) is a plot of the linear restoring force from a silicon spring as
a function of the deformation of the top of the box, assumed to be made of
silicon; \ the solid line (---) is the destabilizing vacuum force on the top
of the box; and the dot-dash line ($-\cdot-$ ) is the total force on the top
of the box. \ Note: The force on the y-axis is actually the total force for
1000 boxes.}%
\label{cavone}%
\end{center}
\end{figure}

The unexpected behavior of forces on the walls of a rectangular cavity
mentioned previously allows us to model a cavity with dimensions such that a
wall vibrates in part due to the vacuum stress\cite{maclaysqueezed}. \ For
example, a cavity that is 2 $\mu m$ long, 0.1 $\mu m$ wide, and about 0.146
$\mu m$ deep will have zero force on the face normal to the 0.146 direction.
\ The zero force corresponds to an unstable energy maximum. \ Thus a
deflection inward leads to an increasing inward (attractive) force, and,
conversely, any deflection outward (repulsive force) leads to an increasing
outward force. \ This potential is akin to a harmonic oscillator, except the
force is destabilizing ($F=kx$) rather than stabilizing ($F=-kx$). \ If we
assume that the box is made of real conductive materials, then there will be a
restoring force due to the material. \ If we include the restoring force that
arises from the small deflection of a deformable membrane as given by Hooke's
Law, then this configuration might become stable if the material force
constant exceeds that for the Casimir force (Fig. \ref{cavone}).%

These results suggest the intriguing possibility of making a structure that
displays simple harmonic motion for small displacements with a frequency that
depends on the difference of the material force constant and the vacuum force
constant. \ The face of a box of the proper dimensions may oscillate under the
mutual influence of the vacuum force and the Young's modulus of the material
(Fig. 4a). \ The oscillations would be damped due to the non-ideal properties
of the material and the friction with the environment (Fig. 4b). \ A zero
point oscillation of the cavity wall would be expected. \ The energy in the
lowest mode would be modified by the temperature.%

\begin{figure}[tbp] \centering
$%
\begin{array}
[c]{cc}%
{\includegraphics[
height=1.772in,
width=2.3359in
]%
{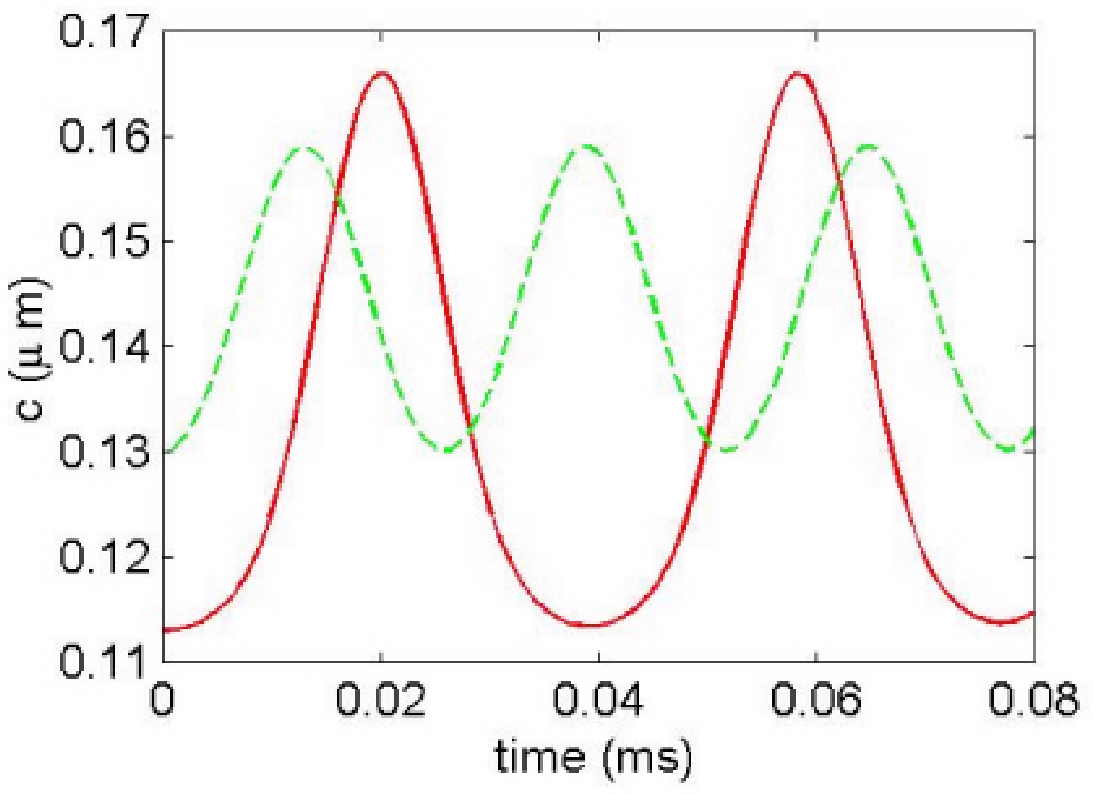}%
}%
&
{\includegraphics[
height=1.7711in,
width=2.3298in
]%
{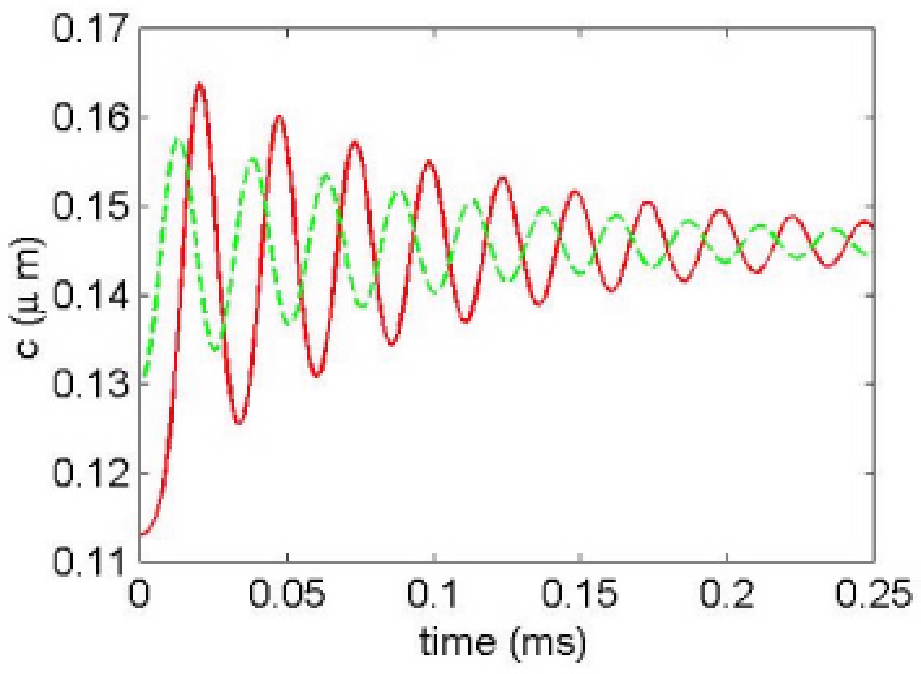}%
}%
\end{array}
$\caption{ a. Displacement of the cover plate as a function of time for two
starting positions. The solid curve is for an initial deflection from the
equilibrium position to a spacing of 0.113 micrometers, close to the minimum for
oscillatory behavior, and clearly shows anharmonic behavior.  The dashed is
for a smaller initial offset from the equilibrium position, and results in a
more sinusoidal motion. b. Displacement vs. time for the same two initial
displacements, but including a damping ratio of 0.025.}%
\end{figure}%

\subsection{Comment on the Exchange of Energy with the Quantum Vacuum}

If QED predicts a large energy density in the quantum vacuum, is there some
way to make use of this vast energy? \ In order to maintain the conservation
of energy, all forms of energy, including vacuum energy, must be included.
\ Thus from the scientific viewpoint, the answer seems clear that it is
possible to transduce vacuum energy into, for example, mechanical energy.
\ However, the process as currently understood does not appear to have any
practical value. \ No one has conceived of a system in which energy can be
extracted in a closed cycle from the vacuum. \ All that can be done is extract
energy in a single operation. To illustrate, consider, an arrangement of two
perfectly conducting, uncharged, parallel plates in a vacuum as an energy
source. \ The Casimir energy $U_{C}(x)$ at zero degrees Kelvin between plates
of area $A$, separated by a distance $x$ is:%
\begin{equation}
U_{C}(x)=-\frac{\pi^{2}}{720}\frac{\hbar cA}{x^{3}}%
\end{equation}

If we allow the plates to move from a large initial separation $a$ to a very
small final separation $b$ then the change in the vacuum energy energy between
the plates is approximately:%
\begin{align}
\Delta U_{C}  &  =U_{C}(b)-U_{C}(a)\\
&  \approx-\frac{\pi^{2}}{720}\frac{\hbar cA}{b^{3}}%
\end{align}

The attractive Casimir force has done work on the plates, and, in principal,
we can build a device to reversibly extract this energy and use it. \ At the
end of the motion ($x=b$), the energy of the electromagnetic field of the
quantum vacuum between the plates has been reduced by the amount of the work
done, so, as is necessary, the total energy is conserved. \ In practice the
closest distance in separation is about a nanometer due to surface roughness.
\ However, in practice, the forces are piconewtons over a distance of
nanometers, so very little useful energy is extracted. \ \ In addition, once
the plates have moved together, and the energy has been extracted, one has to
do the same amount of work to separate the plates and return them to the
initial positions since this is a conservative system\cite{staif00}.

From a more fundamental perspective, utilizing energy of the quantum
fluctuations of the electromagnetic field does not appear to directly violate
known laws of physics according to the work of Forward,  and Cole and
Puthofff, however improbable or impossible such a development might seem at
first consideration \cite{forward}\cite{coleandputhoff}. \ Forward showed that
it is possible to conceive of a device, a foliated capacitor, in which one
could extract electrical energy from the quantum vacuum to do work. The energy
is extracted as the portions of the capacitor that repel each other due to
electrostatic forces come together under the influence of the Casimir
force\cite{forward}. Cole and Puthoff used stochastic electrodynamics to
examine the process of removing energy from the vacuum fluctuations at zero
temperature from the viewpoint of thermodynamics and showed there is no
violation of the laws of thermodynamics\cite{coleandputhoff}. \ \ In the same
spirit, Rueda has suggested that very high energy particles observed in space
may derive their kinetic energy from a long term acceleration due to the
stochastic vacuum field\cite{rueda}. \ In a careful analysis, Cole has shown
that this process of energy transfer from the vacuum field to kinetic energy
of the particles does not violate the laws of thermodynamics\cite{colethermod}%
. \ In stochastic electrodynamics one treats the vacuum fluctuations as a
universal random classical electromagnetic field. \ A formal analogy exists
between stochastic electrodynamics and quantum electrodynamics: \ the field
correlation functions in one theory are related to the Wightman functions in
the other theory\cite{boyer}.

\subsection{Vacuum Forces on Particles}

\ \ As we have mentioned, the parallel plate vacuum forces have been
extensively measured and calculated, and even utilized in a sensitive position
sensor. \ Theoretical forces for geometries with moveable parallel plates, and
rectangular cavities have been explored. \ The question arises: what other
manifestations of vacuum forces, aside from stiction effects, might be of
technological interest as the dimensions of MEMS devices are reduced? \ We
mention a few examples. \ The first has to do with forces on charged or
polarizable particles in the vacuum. \ The electromagnetic field of the ground
state of the quantum vacuum shares the properties of the fields arising from
excited states of the electromagnetic field, when real photons are
present.\ \ Whenever there is an inhomogeneous vacuum energy density, there
will a net force on a polarizable neutral particle given by $\frac{1}{2}%
\alpha\overrightarrow{\nabla}\langle E(x)^{2}\rangle.$ \ Local changes in mode
density and therefore vacuum energy density are induced by the presence of
curved surfaces, and, depending on whether the curvature is positive or
negative, the force between the surface and the particle may be repulsive or
attractive \cite{Deutsch}. \ \ \ The simplest example of a surface altering
the vacuum modes is a perfectly conducting, infinite wall. \ The change in the
vacuum field energy due to the wall produces in this case the well-known
Casimir-Polder interaction: for sufficiently large distances $d$ from the
wall. \ The interaction potential is $V(d)=-3\alpha\hbar c/8\pi d^{4}$, where
$\alpha(0)$ is the static polarizability of the (ground-state) atom. This
effect has been accurately verified in the elegant experiments of Sukenik
\textit{et al} in which he measured the deflection of an atomic beam near a
surface\cite{suk}. \ In this experiment, the particles are actually passed
between the surfaces of a wedge consisting of two conducting planes that
intersect at an angle $\beta$ radians. \ The stress-energy tensor $T^{\mu\nu}$
is not constant in this region, as it is between two parallel plates, but
$T^{00}$ increases as one moves closer to the point of intersection, at which
there is a singularity. \ In the experiment, only the effect of the force
approximately normal to the surfaces was measured. \ As one might expect,
there is also a radial force on a particle at a distance $r$ and at an angle
$\beta/2$ from the intersection that tends to accelerate the particle toward
the intersection provided the static polarizability $\alpha$ is
positive\cite{brevik}:%
\[
F_{r}(r)=-\frac{\alpha(0)\hslash c}{90\pi r^{5}\beta^{4}}(44\pi^{4}+80\pi
^{2}\beta^{2}+11\beta^{4})
\]
For the case of $\beta=\pi$ we have a particle near a plane and recover the
usual Casimir-Polder force. \ The tangential force in the $\theta$ direction
vanishes along this midline. \ Note that there would be a torque on a
permanent dipole in this wedge.

There are many geometries for which the stress-energy tensor has not been
computed as a function of position, and we do not know what the forces on a
charged or polarizable particle in the vacuum might be. \ Consider for
example, the forces on a particle within a closed rectangular cavity, where
the kinetic energy of the particle is much less than the change in vacuum
energy due to the surfaces. \ Very near any surface, away from edges and other
walls, one might expect the particle to experience the usual Casimir-Polder
force. \ In other regions of the cavity, the forces are not know since
calculations of the stress-energy tensor have not been done without averaging
over the entire volume. \ What is the equilibrium state of a group of atoms or
particles in a region of altered vacuum energy? \ For example, assume we have
a number of particles in a metal sphere or a metal box in which the vacuum
modes have been altered from the free field modes. \ What is the equilibrium
distribution of these particles? \ Since there is a non-homogeneous vacuum
field, the particles will experience forces. \ Will there be some vacuum
damping that gives them a terminal velocity? \ Will the particles congregate
in a region of the lowest energy? \ Will they bounce off the walls and give
some kind of force on the walls. Are these vacuum forces negligible, except at
very low temperatures? \ The motion of one particle inside a box or sphere
would be interesting. \ Does the interaction provide a window into vacuum
energy so that we can make two reservoirs to operate an engine?? \ If a hole
is put in one of the sides of the box, what happens? \ There is one
calculation that suggests that very high energy particles observed in space
may derive their kinetic energy from a long term acceleration due to the
stochastic vacuum field\cite{rueda}.

\subsection{Systems with Torques}

Consider the conditions for which we would expect a medium, such as a
dielectric slab, to experience a torque in the vacuum. \ If we view the origin
of a vacuum torque as the transfer of the angular momentum of zero-point
photons to the medium, then it is clear that\ to have a torque we need to have
a geometrical configuration in which the vacuum energy depends on the angular
orientation of the medium. \ This requirement cannot be met with a single
object, even if it is not isotropic. \ However, two plates separated by a
distance $d$ that are birefringent would break the rotational symmetry of the
vacuum and be expected to experience a torque. This torque has indeed been
calculated, and compared to the attractive Casimir force between the
plates\cite{enk}. \ \ From dimensional grounds the torque\ between two thick
plates (thickness%
$>$%
$>$%
separation) of area $A$ goes as $f\hslash cA/d^{3},$ where Enk has derived an
expression for the dimensionless number $f$ which is determined by the square
of the difference of the refractive indices, and has a typical value of about
$10^{-6}.$ \ The torque, which appears measurable, varies as $sin2\phi$, where
$\phi$ is the angle between the two optic axes. \ The dielectrics tend to
rotate in opposite directions so the total angular momentum transfer from the
vacuum is zero.

\subsection{Forces on Semiconductor Surfaces}

One of the potentially most important configurations from the technological
viewpoint involves vacuum forces on semiconductor surfaces. \ The Casimir
force for a conducting material depends on the plasma frequency, beyond which
the material tends to act like a transparent medium. \ For parallel plates
separated by a distance $d$ the usual Casimir force is reduced by a factor of
approximately $C(a)=(1+(8\lambda_{p}/3\pi d))^{-1},$ where $\lambda_{p}$ is
the wavelength corresponding to the plasma frequency of the
material\cite{lambrecht}. \ Since the plasma frequency is proportional to the
carrier density, it is possible to tune the plasma frequency in a
semiconductor, for example, by illumination or by temperature, or by the
application of a voltage bias. \ In principle it should be possible to build a
Casimir switch that is activated by light, a device that would be useful in
optical switching systems. \ A very interesting measurement of the Casimir
force between a flat surface of borosilicate glass and a surface covered with
a film of amorphous silicon was done in 1979 by Arnold et al\cite{arnold}
\ They observed an increase in the Casimir force when the semiconductor was
exposed to light. \ This experiment has yet to be repeated with modern methods
and materials. \ As a first step, Chen et al have measured used an AFM to
measure the force between a single Si crystal and a 200 $\mu m$ diameter gold
coated sphere, and found good agreement with theory using the Lizshitz
formalism\cite{chensilicon}.

\subsection{Vacuum Powered Space Craft}

It is possible, albeit impractical, to conceive of a vacuum spacecraft that
operates by pushing on the quantum vacuum\cite{maclayandforward}. \ With a
suitable trajectory, the motion of a mirror in vacuum can excite the quantized
vacuum electromagnetic field with the creation of real photons. \ This
possibility was first noticed in 1970, when Moore considered the effect of an
uncharged one dimensional boundary surface in vacuum that moved, with the very
interesting prediction that it should be possible to generate real photons
from a suitable motion\cite{moore}\cite{law}. \ This effect, referred to as
the dynamic or adiabatic Casimir effect, has been reviewed but not verified
experimentally \cite{plunienreview}\cite{bordag}\cite{birrellanddavies}.
\ Energy conservation requires the existence of a radiation reaction force
working against the motion of the mirror \cite{netoandmachado}, and this force
can result in a net acceleration of the mirror. \ The vacuum field exerts a
force on the moving mirror that tends to damp the motion. \ This dissipative
force may be understood as the mechanical effect of the emission of radiation
induced by the motion of the mirror. \ The energy expended moving the mirror
against the radiative force goes into electromagnetic radiation.

The Casimir drive spacecraft is not suggested as practical way to build a
spacecraft, but to illustrate another potential role of the quantum vacuum.
\ Perhaps a more clever quantum drive will some day become practical or other
uses of the dynamic Casimir effect will arise. \ Physicists have explored
various means of locomotion depending on the density of the medium and the
size of the moving object. \ It would be interesting to find an optimum method
for moving in the quantum vacuum. \ Unfortunately we currently have no simple
way to mathematically explore various simple possibilities.

\section{Conclusion}

The are many potential ways in which the ground state of the vacuum
electromagnetic field might be engineered for use in technological
applications, a few of which we have mentioned here. \ As the technology to
fabricate small devices improves, as the theoretical capability of calculating
quantum vacuum effects increases, it will be interesting to see which
possibilities prove to be useful and which just remain curiosities, and which
limit the performance of MEMS. \ In a way the situation is reminiscent of
electricity in the 1600s, when Faraday was asked of what use is electricity?,
and answered "Of what use is a new born baby?". \ We are not very good at
predicting the development of technology. \ In the 1980's many thought AI
would revolutionize the world, but it didn't. \ In the 1960s, manufacturers
were hard put to think of any reason why an individual would want a home
computer and today we wonder how we ever survived without them.

\begin{acknowledgments}
We would like to thank the NASA Breakthrough Propulsion Program for their
support, and Jay Hammer for finite element calculations of the vibrating
cavity walls.
\end{acknowledgments}

\end{document}